\documentstyle[preprint,prd,aps]{revtex}
\begin{document}
\draft
\title{Tensor-Induced CMB Temperature-Polarization \\
Correlation in Reionized Universes}
\author{Kin-Wang Ng\footnote{\tt nkw@phys.sinica.edu.tw}} 
\address{Institute of Physics \& Institute of Astronomy and Astrophysics\\
Academia Sinica, Taipei, Taiwan 11529, R.O.C.}
\maketitle

\begin{abstract}
We reexamine the temperature-polarization correlation function of the
cosmic microwave background induced by tensor mode with a scale-invariant 
spectrum in reionized standard cold dark matter models. 
It is found that the sign of the correlation function is positive 
on all angular scales even in a model with substantial reionization.
\end{abstract}
\bigskip

The detection of the large-angle anisotropy of the cosmic microwave
background (CMB) by the {\it COBE} DMR experiment~\cite{smo} has incited
a lot of studies in this area~\cite{tur}.
It is now well established that CMB anisotropies are genuine 
imprint of the early universe, which could potentially be used to determine
to a high precision virtually all cosmological parameters of
interest~\cite{jun}. Future missions such as NASA MAP
and ESA Planck Surveyor would measure the anisotropy spectrum 
in high-precision;
the polarization spectrum, expected with an order of magnitude below the
anisotropy, would also be measured to provide complementary information to
anisotropy.

It was first pointed out by Crittenden {\it et al.} that the correlation
between the temperature and polarization anisotropies offers a test of
physics on the last scattering surface, as well as a possibility of
distinguishing the scalar and tensor perturbations~\cite{cri1,cri2}. 
They found that in a universe with standard recombination
the sign of the tensor-induced temperature-polarization (TQ) 
correlation function is opposite to that for scalar mode on large scales. 
With substantial early ionization, 
polarization is greatly enhanced on large scales and
a geometrical effect causes the tensor correlation function to reverse 
sign for $\theta > 30^o$~\cite{cri2}. 

TQ correlation will be 
measured by MAP and Planck at small scales. In addition, polarization 
experiments such as ground-based POLAR~\cite{kea} and SPOrt/ISSA~\cite{spo} 
will observe large-scale polarization as to test the thermal history of the
early universe. In particular, the latter with full-sky coverage will
be sensitive to TQ correlation~\cite{ng5}. 
Although TQ correlation has been studied extensively~\cite{hu1,hu2},  
it is worthwhile to reexamine and work out explicitly 
the TQ correlation function in reionized universes. 
We will follow closely the method of Ref.~\cite{cri1,cri2}.
We find a correction term to their tensor correlation function
that renders it a positive function irrespective of the reionization history.
Our result should be useful to the large-scale polarization experiments.

Here we do not repeat detailed calculations. Only necessary steps 
for obtaining correlation functions are given.
For tensor $+$-mode, the Stokes parameters induced by a Fourier mode with 
wavevector ${\bf k}$ in the direction of an unit vector $\hat{\bf p}$ are:
\begin{equation}
\left( \begin{array}{c} T'_{\bf k}\\Q'_{\bf k}\\U'_{\bf k} \end{array} \right)=
\left( \begin{array}{c}
{1\over2}\alpha(\mu)(1-\mu^2)\cos2\phi\\ 
{1\over2}\beta(\mu)(1+\mu^2)\cos2\phi\\ 
\beta(\mu)\mu\sin2\phi
\end{array} \right),
\label{tqu}
\end{equation} 
where $\mu \equiv \hat{\bf p}\cdot\hat{\bf k}$ and $\phi$ is the 
azimuthal angle of $\hat{\bf p}$ about $\hat{\bf k}$.
The $\times$-mode solution is given by the same expressions, 
except for replacing $\cos2\phi$ by $\sin2\phi$, 
and $\sin2\phi$ by $-\cos2\phi$.
For scalar mode, $T'_{\bf k}=\alpha(\mu)$, $Q'_{\bf k}=\beta(\mu)$,
and $U'_{\bf k}=0$. 
Expanded in Legendre polynomials,
$\alpha(\mu) = \sum _{l} (2l+1) \alpha_l P_l(\mu)$ and 
$\beta(\mu)=\sum _{l} (2l+1) \beta_l P_l(\mu)$.
We evaluated all $\alpha_l$'s and $\beta_l$'s using
the Boltzmann numerical code developed in Ref.~\cite{ng23}.

To obtain the Stokes parameters defined with respect to a 
fixed orthonormal basis $({\bf e}_x,{\bf e}_y,{\bf e}_z)$
from those defined in the $\hat{\bf k}$-basis~(\ref{tqu}), 
we need to perform the rotation~\cite{bon}:
\begin{equation}
\left(\begin{array}{c} T_{\bf k}\\Q_{\bf k}\\U_{\bf k} \end{array}\right)=
\left(\begin{array}{ccc} 
1 & 0 & 0 \\
0 & \cos2\psi & \sin2\psi \\
0 & -\sin2\psi & \cos2\psi
\end{array}\right)
\left(\begin{array}{c} T'_{\bf k}\\Q'_{\bf k}\\U'_{\bf k} \end{array}\right),
\end{equation}
where the rotation angle $\psi$ is given by
\begin{eqnarray}
\cos\psi&=&
\frac{\hat{\bf k}\cdot{\bf e}_z-(\hat{\bf k}\cdot\hat{\bf p})
({\bf e}_z\cdot\hat{\bf p})}{
\sqrt{1-(\hat{\bf k}\cdot\hat{\bf p})^2} 
\sqrt{1-({\bf e}_z\cdot\hat{\bf p})^2}}, \nonumber \\
\sin\psi&=&
\frac{\hat{\bf p}\cdot({\bf e}_z \times \hat{\bf k})}{
\sqrt{1-(\hat{\bf k}\cdot\hat{\bf p})^2} 
\sqrt{1-({\bf e}_z\cdot\hat{\bf p})^2}}.
\label{rot}
\end{eqnarray}

Summing up all ${\bf k}$-mode contributions and two polarizations,
the tensor TQ correlation function is obtained as
\begin{eqnarray}
\langle T(\hat{\bf p}_1) Q(\hat{\bf p}_2) \rangle =
{1\over 4}\int d^3 {\bf k} && \left[(1-\mu_1^2)(1+\mu_2^2) 
\cos2(\phi_1-\phi_2) \cos2\psi_2 \right. \nonumber \\
&& \left.-2(1-\mu_1^2)\mu_2 \sin2(\phi_1-\phi_2) \sin2\psi_2 \right]
\alpha(\mu_1)\beta(\mu_2),
\label{ttq1}
\end{eqnarray}
where $\mu_1=\hat{\bf p}_1 \cdot \hat{\bf k}$
and $\mu_2=\hat{\bf p}_2 \cdot \hat{\bf k}$, $\phi_1$ and $\phi_2$
are respectively the azimuthal angles of $\hat{\bf p}_1$ and $\hat{\bf p}_2$
about $\hat{\bf k}$, and $\psi_2$ is given by Eq.~(\ref{rot}) with 
$\hat{\bf p}$ replaced by $\hat{\bf p}_2$.
It is straightforward to prove the identities,
\begin{eqnarray}
\cos(\phi_1-\phi_2)&=&
\frac{\hat{\bf p}_1\cdot\hat{\bf p}_2-\mu_1 \mu_2}{
\sqrt{1-\mu_1^2} \sqrt{1-\mu_2^2}}, \nonumber \\
\sin(\phi_1-\phi_2)&=&
\frac{(\hat{\bf p}_2 \times \hat{\bf p}_1)\cdot \hat{\bf k}}{
\sqrt{1-\mu_1^2} \sqrt{1-\mu_2^2}}.
\end{eqnarray}

Without loss of generality, we choose $\hat{\bf p}_1=\hat{\bf q}$ and
$\hat{\bf p}_2={\bf e}_z$, and use the axes ${\bf e}_x$ and ${\bf e}_y$
to define the Stokes parameter $Q({\bf e}_z)$. 
Hence, $\mu_2=\mu_{\hat{\bf k}}$ and $\psi_2=\phi_{\hat{\bf k}}$,
where $(\mu_{\hat{\bf k}}, \phi_{\hat{\bf k}})$ are the spherical polar 
angles of $\hat{\bf k}$ in the coordinate $({\bf e}_x,{\bf e}_y,{\bf e}_z)$.
Then, expanding
\begin{eqnarray}
A(\mu_1)&\equiv&\mu_1^2\alpha(\mu_1)=\sum _{l} (2l+1) A_l P_l(\mu_1),
\nonumber \\
B(\mu_2)&\equiv&(1+\mu_2^2)\beta(\mu_2)=\sum _{l} (2l+1) B_l P_l(\mu_2),
\end{eqnarray}
and integrating over $\mu_{\hat{\bf k}}$ and $\phi_{\hat{\bf k}}$, we obtain 
\begin{eqnarray}
\langle T(\hat{\bf q}) Q({\bf e}_z) \rangle &=& {\pi\over2} \cos 2\varphi 
\int k^2dk \sum_{l,l'} (2l+1)(2l'+1) \nonumber \\ 
&&\times \biggl[ {(l'-2)!\over (l'+2)!}
[\alpha_{l'} B_l \cos^2\theta-A_{l'} B_l] a^2_{ll'} P^2_{l'}(\cos\theta) 
\nonumber \\
&& + \frac{1}{2}\sin^2\theta \alpha_{l'} B_l \biggl( \delta_{ll'}
\frac{2}{2l+1} P_{l'}(\cos\theta)
+{(l'-4)!\over (l'+4)!} a^4_{ll'} P^4_{l'}(\cos\theta) \biggr) \nonumber \\
&& -\sin2\theta \alpha_{l'} \beta_l \biggl( {(l'-1)!\over (l'+1)!} 
b^1_{ll'} P^1_{l'}(\cos\theta)
-{(l'-3)!\over (l'+3)!} b^3_{ll'} P^3_{l'}(\cos\theta) \biggr) \nonumber \\
&& +\sin^2\theta \alpha_{l'} (B_l-\beta_l)
\biggl( \delta_{ll'}\frac{2}{2l+1} P_{l'}(\cos\theta)
-{(l'-4)!\over (l'+4)!} a^4_{ll'} P^4_{l'}(\cos\theta) \biggr) \biggr],
\label{ttq2}
\end{eqnarray}
where $(\theta, \varphi)$ are the spherical polar angles
of $\hat{\bf q}$ about ${\bf e}_z$.
The constants $a^n_{ll'}$ and $b^m_{ll'}$ are given respectively by
$a^n_{ll'}=\int_{-1}^1dx P_{l}(x) P_{l'}^n(x)$ and
$b^m_{ll'}=\int_{-1}^1dx x(1-x^2)^{1\over2} P_l(x) P_{l'}^m(x)$.
In Eq.~(\ref{ttq2}), the first two terms are the expression for the tensor
$\langle TQ \rangle$ that was found in Ref.~\cite{cri2}. 
They come from integration of the first term containing $\cos2\psi_2$ 
in Eq.~(\ref{ttq1}).
The remaining terms are from integrating the $\sin2\psi_2$ term.
Fig.~1 is the plot of $\langle TQ \rangle$ from numerical 
calculations of Eq.~(\ref{ttq2}) with $\varphi=0$ 
for reionized models with different optical depths.
Apparently, the tensor $\langle TQ \rangle$ functions
are positive on all angular scales. The dashed line is the case
of substantial reionization without the correction term,
which is similar to the result obtained in Ref.~\cite{cri2}. 
We can see that the correction is sizable on large angular scales. 
This is expected because the effect of basis rotation operates on 
only large angular scales~\cite{ng4}. 
Thus, in calculating small-scale correlation,
one can simply use the small-angle approximation by making $\psi_2=0$, 
under which $\sin2\psi_2=0$ and the correction term is then switched off.  
In the case of standard recombination, since the large-scale polarization
is extremely small, the correction is not apparent at all.
  
To compare with the scalar mode, in Fig.~2 
we plot the scalar $\langle TQ \rangle$~\cite{cri1,ng4,mel},
\begin{eqnarray}
&&\langle T(\hat{\bf q}) Q({{\bf e}_z}) \rangle
=2\pi\cos2\varphi \sum_{l\ge 2} (2l+1) C_l^{TQ} P_l^2(\cos\theta), 
\nonumber \\
&& C_l^{TQ} = \frac{(l-2)!}{(l+2)!} \sum_{l'}(2l'+1) a^2_{l'l}
\int k^2 dk \alpha_l^* \beta_{l'}.
\label{stq}
\end{eqnarray}
Note that the peak height of scalar $\langle TQ \rangle$ 
in the case of substantial reionization is significantly less than that in
Ref.~\cite{cri1}. This is due to the fact that 
they used a slightly different constant $a^2_{ll'}$ (transposing $l'$ and $l$)
for $C_l^{TQ}$ in the correlation function.
Figs.~1 and~2 show that 
on large scales the tensor $\langle TQ \rangle$ has correlation 
whereas the scalar $\langle TQ \rangle$ has anticorrelation.
In fact this can be easily explained by solving the collisional 
Boltzmann equation in the long-wavelength limit~\cite{ng1}. 
It has been pointed out that in an open universe the sign of the scalar 
$\langle TQ \rangle$ on the largest scales is reversed, however, 
this effect will be destroyed by even
minimal amounts of reionization~\cite{hu2}. 
Thus the positivity of the large-scale TQ correlation is a direct indicator
of a significant tensor component in metric fluctuations.   

More recently, calculations of correlation functions were usually performed
by using an elegant formalism developed by
Zaldarriaga {\it et~al.}~\cite{zal} and Kamionkowski {\it et~al.}~\cite{kam}.
Basically, the method is to expand $Q$ and $U$ in terms of spin-2 
spherical harmonics. In Ref.~\cite{kam}, they gave 
\begin{equation}
\langle T(\hat{\bf q}) Q({\bf e}_z) \rangle 
= -\cos2\varphi \sum_l\frac{2l+1}{4\pi}\sqrt{\frac{(l-2)!}{(l+2)!}}C_{Cl}
   P^2_l(\cos\theta),
\label{cmbfast}
\end{equation}
where the angular spectrum $C_{Cl}$ can be expressed in terms of
$\alpha_l$ and $\beta_l$. For scalar mode, their $\langle TQ \rangle$
is exactly the same as Eq.~(\ref{stq}). 
But their tensor $\langle TQ \rangle$ has a different form 
from Eq.~(\ref{ttq2}), where they can obtain a much simpler 
expression for $C_{Cl}$. Since their expansion scheme is different from ours, 
it is rather difficult to show the equivalence analytically.
However, we compare our numerical results of Eq.~(\ref{ttq2}) with those 
using Eq.~(\ref{cmbfast}) with $C_{Cl}$'s evaluated by using the 
CMBFAST code~\cite{sel}, and good agreements are found.
 
The author would like to thank G.-C. Liu for his numerical work. 
This work was supported in part by the R.O.C. NSC Grant No.
NSC88-2112-M-001-042.

\newpage

\begin{center}
{\bf FIGURE CAPTIONS}
\end{center}
\bigskip
\noindent
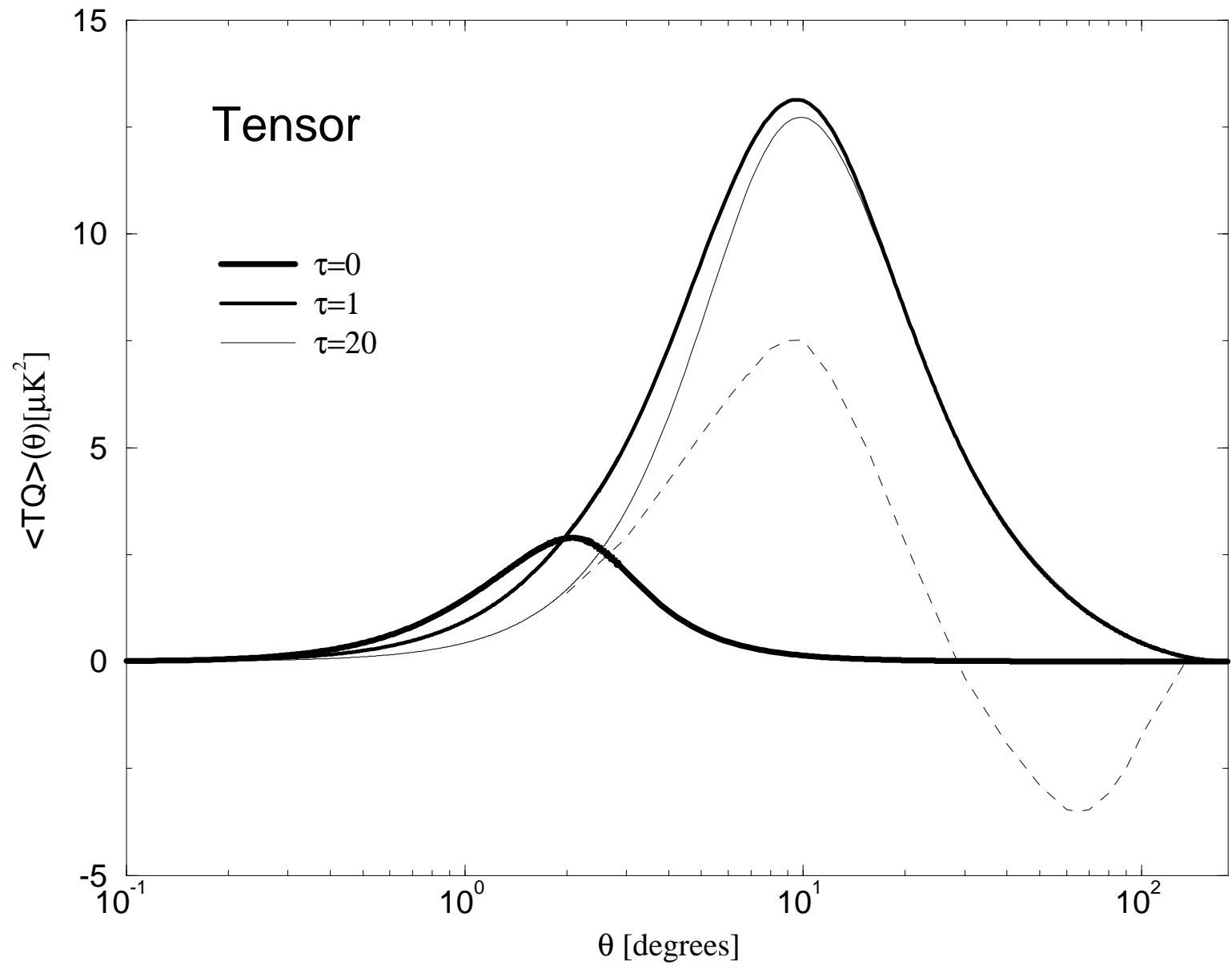
Fig.1. Temperature-polarization correlation functions for scale-invariant 
tensor perturbation in reionized cold dark matter
models ($\Omega_0=1, h=0.5, \Omega_B=0.05$) with different 
reionization optical depths $\tau$. 
Dashed line is for the case of $\tau=20$ without the correction term.
All curves are normalized to COBE data.

\medskip
\noindent
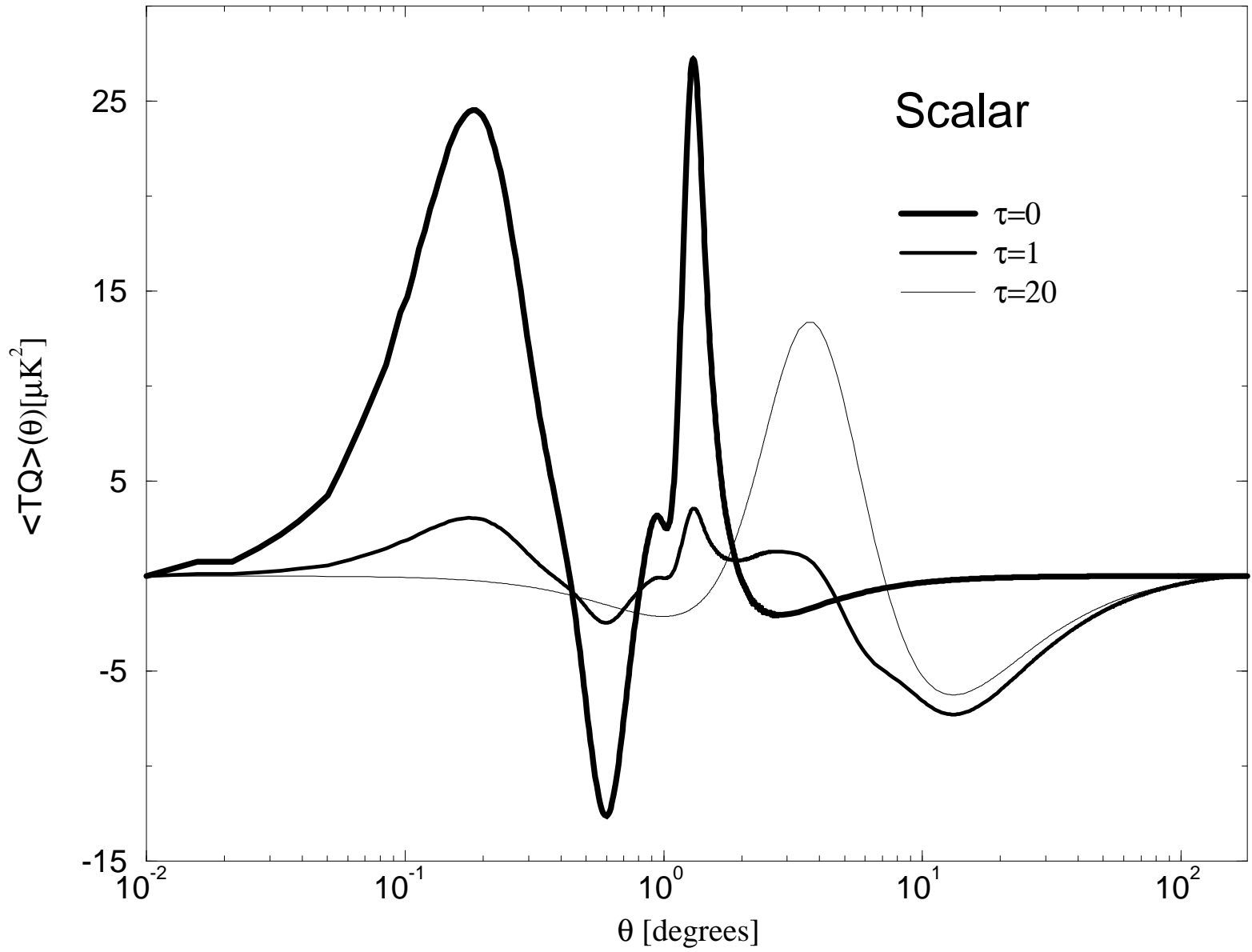
Fig.2. As Fig.~1, but for scale-invariant scalar perturbation. The height
of the peak near $\theta\simeq 4^o$ is about a factor of $3$ less than
that in Ref.~\cite{cri1}.


\begin{references}
\bibitem{smo}
G. F. Smoot {\it et al.}, Astrophys. J. {\bf 369}, L1 (1992).
\bibitem{tur}
For a review, see {\it Critical Dialogues in Cosmology}, edited by N. Turok
(World Scientific, 1997) p.343-395.
\bibitem{jun}
G. Jungman, M. Kamionkowski, A. Kosowsky, and D. N. Spergel, Phys. Rev. D
{\bf 54}, 1332 (1996);
M. Zaldarriaga, D. N. Spergel, and U. Seljak, Astrophys. J. {\bf 488}, 1 (1997).
\bibitem{cri1}
D. Coulson, R. G. Crittenden, and N. G. Turok, Phys. Rev. Lett. {\bf 73},
2390 (1994).
\bibitem{cri2}
R. G. Crittenden, D. Coulson, and N. G. Turok, Phys. Rev. D {\bf 52}, 
R5402 (1995).
\bibitem{kea}
B. Keating, P. Timbie, A. Polnarev, and J. Steinberger, Astrophys. J. 
{\bf 495}, 580 (1998).
\bibitem{spo}
See the SPOrt homepage: http://tonno.tesre.bo.cnr.it/~sport/.
\bibitem{ng5}
K.-W. Ng and G.-C. Liu, Int. J. Mod. Phys. D {\bf 8}, 61 (1999).
\bibitem{hu1}
W. Hu and M. White, New Astronomy {\bf 2}, 323 (1997); and references therein.
\bibitem{hu2}
W. Hu, U. Seljak, M. White, and M. Zaldarriaga, Phys. Rev. D {\bf 57}, 
3290 (1998).
\bibitem{ng23}
K. L. Ng and K.-W. Ng, Astrophys. J. {\bf 445}, 521 (1995);
Astrophys. J. {\bf 456}, 413 (1996).
\bibitem{bon}
J. R. Bond and G. Efstathiou, Mon. Not. R. Astr. Soc. {\bf 226}, 655 (1987).
\bibitem{ng4}
K. L. Ng and K.-W. Ng, Astrophys. J. {\bf 473}, 573 (1996).
\bibitem{mel}
A. Melchiorri and N. Vittorio, Proc. of NATO Advanced Study Institute 1996,
astro-ph/9610029.
\bibitem{ng1}
K. L. Ng and K.-W. Ng, Phys. Rev. D {\bf 51}, 364 (1995);
D. Harari and M. Zaldarriaga, Phys. Lett. B {\bf 319}, 96 (1993).
\bibitem{zal}
M. Zaldarriaga and U. Seljak, Phys. Rev. D {\bf 55}, 1830 (1997).
\bibitem{kam}
M. Kamionkowski, A. Kosowsky, and A. Stebbins, Phys. Rev. D {\bf 55},
7368 (1997).
\bibitem{sel}
U. Seljak and M. Zaldarriaga, Astrophys. J. {\bf 469}, 437 (1996).
\end{references}
\end{document}